\documentclass[]{iopart}
\usepackage{iopams}  
\usepackage{graphicx, epsfig}
\begin{document}

\def\be{\begin{equation}}
\def\ee{\end{equation}}
\def\bea{\begin{eqnarray}}
\def\eea{\end{eqnarray}}

\title[]{ Real space information from Fluctuation electron microscopy: Applications to amorphous 
silicon
}

\author{Parthapratim Biswas}
\address{Department of Physics and Astronomy, The University of Southern Mississippi, 
Hattiesburg, MS 39401, USA} 
\ead{partha.biswas@usm.edu} 

\author{Raymond Atta-Fynn}
\address{Department of Physics and Astronomy, The University of Texas, Arlington, TX 76019, USA} 
\ead{attafynn@uta.edu}

\author{S. Chakraborty and D. A. Drabold}
\address{Department of Physics and Astronomy, Ohio University, Athens, OH 45701, USA} 
\ead{drabold@ohio.edu}

\begin{abstract} 

Ideal models of complex materials must satisfy all available information about the system. 
Generally, this information consists of experimental data, information implicit to sophisticated 
interatomic interactions and potentially other {\it a priori} information. 
By jointly imposing first-principles or tight-binding information in 
conjunction with experimental data, we have developed a method:  Experimentally Constrained 
Molecular Relaxation (ECMR) that uses {\it all} of the information available. We apply the 
method to model medium range order in amorphous silicon using Fluctuation Electron microscopy 
(FEM) data as experimental information. The paracrystalline model of medium range order is 
examined,  and a new model based on voids in amorphous silicon is proposed. Our work suggests 
that films of amorphous silicon showing medium range order (in FEM experiments) can be accurately 
represented by a continuous random network model with inhomogeneities consisting of ordered 
grains and voids dispersed in the network. 

\end{abstract}




\section{Introduction}

The structural modeling of amorphous materials poses a particular challenge to condensed 
matter science. The initial hurdle to overcome is devising a computer model that accurately 
represents a small fragment of the material. Experimental data is inevitably the result of 
a system average involving macroscopic number of atoms in a continuously variable range of 
conformations. The result is that such data tend to be smooth with very limited information 
content. While the information provided by experiments is evidently of critical importance 
to understanding these materials,  such information is incomplete (e.g., the information in 
the data is incapable of uniquely specifying the structure). The impressive advances in protein 
crystallography help to illustrate the challenge: in any crystalline system, diffraction 
measurements yield a palisade of $\delta$ functions. From the information entropy~\cite{jaynes} 
it is easy to show that there is vastly more information in the sharply defined function for 
the crystal than the smooth function characteristic of a glass or amorphous material. The 
structure factor for the crystal is nearly sufficient to uniquely invert the data to obtain 
the structure, a stark contrast with the situation for amorphous materials. This argument 
also emphasizes the need to use {\it all} available experiments in modeling. Despite our 
lamentations about the limitations of information-based modeling, it is clearly wise to 
build models consistent with experimental information: our concern is that this information 
is highly incomplete by itself.  

The limitations of information from experimental data beg for a molecular dynamics (MD) or 
Monte Carlo modeling approach using accurate interatomic interactions. If properly implemented, 
such a scheme will enforce the proper local ordering, chemistry etc. However, these approaches 
suffer from their own shortcomings: despite superficial similarities to the physical process 
of making a glass (quenching from the melt),  such simulations are carried out with unphysically 
rapid quenches, models that are tiny (especially if accurate interactions are used), and of 
course the interactions themselves are never perfect. Despite these cautions, such simulations 
have met with many successes in a range of materials.

An ideal modeling approach should merge the information-based method and the computer simulation scheme. There 
is no unique way to accomplish this, and the ``bottom line" is that whatever scheme is adopted, it must produce 
models that agree with all known information. We are aware of three efforts in this direction: our Experimentally 
Constrained Molecular Relaxation" (ECMR) method~\cite{ecmr}, a Bayesian method for biomolecules~\cite{pnas} and 
a related scheme used on amorphous carbon~\cite{opletal}. These methods vary in many details, but are similar 
in spirit and all have met with success in the problems approached.  

Hydrogenated a-Si (a-Si:H) is one of the most important electronic materials~\cite{street}.  While there is 
slight variability in pair-correlation functions measured for different samples, Fluctuation Electron 
Microscopy (FEM) experiments probing triplet or higher atomic correlations show dramatic variation from 
sample to sample. Even in this most venerable amorphous electronic material there is a lack of understanding 
about the difference in network topology on the medium range length scale between samples with different 
FEM data.  In this paper we further develop our ECMR method to form models of a-Si including medium range 
order implied by Fluctuation Electron Microscopy (FEM) measurements.

\section{The Inverse problems in materials modeling }

The inverse approach takes a very different route to model materials. The focus here is on 
available experimental information pertaining to the materials under study. The challenge is to 
construct a model that is consistent with a given set of experimental data, and additionally 
an approximate total energy functional.  In the context of materials modeling, the primary interest is on  structure determination 
and the resulting electronic properties, but the formalism is also useful to construct 
empirical potentials~\cite{Lyubartsev, Soper}. Although there exists no general proof that 
a many-body potential can be constructed uniquely within this approach, Henderson 
has shown a connection between pair potentials and radial distributions that states for a 
system under given temperature and pressure two pair potentials that produce same radial 
distribution functions can differ only by an additive constant~\cite{Henderson}. Lyubartsev 
and Laaksonen have followed this idea to construct interaction potentials from radial distribution 
functions via reverse Monte Carlo simulation and apply it to aqueous sodium chloride (NaCl) 
solution~\cite{Lyubartsev}. Soper has developed empirical potential structure refinement (EPSR) 
where total diffraction data can be inverted into a set of partial structure factors by 
extending an earlier method of Edwards and Enderby~\cite{Edwards} and reverse Monte Carlo 
method~\cite{Soper}.  Zunger has recently applied the inverse band structure approach to 
find atomic configurations for a given set of electronic and optical properties in alloys~\cite{Zunger}. 

The reverse Monte Carlo (RMC) method developed by McGreevy and coworkers describes how to construct a physical 
structure (i.e. a 3-dimensional model) of a material using the information included in the structure 
factors~\cite{McG, Gere, Walters, Biswas}.  Instead of using any conventional energy functional, a generalised 
penalty function is constructed involving experimental structural data and some suitable constraints, 
which is then minimized by using the Metropolis Monte Carlo algorithm~\cite{mc}. The set of configurations 
obtained in this method can be used for further analysis of structural, electronic and vibrational 
properties. The method does not generate interaction potentials and in absence of sufficient information, 
configurations obtained from RMC may not be physically meaningful. One usually addresses this problem 
by adding further information, but often this proves to be difficult to optimize via 
simple Monte Carlo scheme. 

ECMR has been designed to overcome some of the problems above~\cite{ecmr}. Mathematically, ECMR 
offers an approximate solution to the constrained optimization problem: {\it Find a set of coordinates 
that is a minimum of an accurate energy functional subject to the constraint that the coordinates 
reproduce one or more experimental data sets}. In practice it may be useful to impose other constraints 
too, for example on atomic coordination or chemical order.  In the following, we apply ECMR to model 
medium range order using FEM data as experimental information and an empirical total energy functional.

\section{Modeling medium range order as an inverse problem} 
Medium range order (MRO) is defined as structural ordering that exists between the short 
range (typically 3-5 {\AA}) and the long range ($> 300$ {\AA}) length scale~\cite{elliott-book}. 
Quantifying order at this length scale is somewhat ambiguous and requires information 
beyond radial (pair) distribution functions. Until recently, there has been a very few 
direct experimental evidence to detect MRO. 
In ionic and covalent glasses, MRO manifests itself 
in the first sharp diffraction peak (FSDP) of the total factor structure factor~\cite{Elliott1}. 
This feature corresponds to real space ordering in materials at the intermediate length 
scale.  The well known Staebler-Wronski effect is an example where creation of metastable dangling 
bonds in hydrogenated amorphous silicon upon exposure to visible light~\cite{Staebler} 
has been observed to occur in the material with diminishing medium range order~\cite{Gibson1}. 
Fluctuation electron microscopy clearly reveals that structure of thin films of amorphous 
silicon are much more complex than a continuous random network model~\cite{Voyles}. 

Higher order correlation functions are the most suitable candidates for studying the signature 
of MRO in amorphous networks. However, obtaining experimental 
structural information beyond the 2-body correlation function is 
non-trivial and there exists no simple and direct scheme of systematic analysis of the full 3- 
and 4-body correlation functions. Treacy and Gibson have addressed the problem experimentally 
by developing a low resolution electron microscopy technique known as fluctuation electron 
microscopy (FEM)~\cite{fem1}. FEM can detect MRO because it is sensitive to 3- and 4-body correlation 
functions. 
It was shown that the fluctuation in the diffracted intensities can be measured by the normalized variance of the 
intensities, and is directly related to 3- and 4-body correlation functions containing 
the information at the medium range length scale~\cite{Voyles}. 

We apply our ECMR technique starting with two very different models of a-Si: the first is a paracrystalline model 
of amorphous Si proposed by  Khare~\cite{Khare} and the second includes voids in continuous random networks.  In 
our work, we start from each of these models and apply our ECMR method to obtain final configurations
displaying FEM signal, which we call Model-A and Model-B respectively.  In both the cases, one 
observes the presence of strong FEM signal, and the model is also consistent with other physical 
observables such as structure factors, electronic and vibrational density of states.

\section{Paracrystalline models of medium range order } 

Before we proceed to model generation, we briefly mention the key equations of Fluctuation Electron 
Microscopy (FEM) that have been used here in conjunction with ECMR method to generate amorphous network containing 
medium range order. For a detailed description of FEM and ECMR, we refer to Refs ~\cite{Voyles, fem1, fem2} and 
Ref~\cite{ecmr} respectively.  In FEM, we estimate MRO by measuring the normalized variance of the dark-field image intensity instead 
of intensity itself. The normalized variance is defined as: 

\be 
V(k, Q)=\frac{\langle{I^{2}( k, Q)}\rangle} {{\langle{I(k, Q)}\rangle}^{2}} - 1 
\label{vk}
\ee

The variable ${\bf k}$ is the magnitude of the scattering vector and $1/Q$ defines the 
characteristic length scale of MRO. In a variable coherence microscopy, one 
fixes the value of $Q$ and varies ${\bf k}$ in order to determine the degree of MRO present in 
the length scale of inverse $Q$.  Following Treacy and Gibson~\cite{Gibson1, fem1}, 
we are interested in the fluctuation in the intensity for varying ${\bf k}$ at a 
fixed spatial resolution.  The intensity $I({\bf k},Q)$ due to scattering from a 
volume centered at {\bf r} of size proportional to $1/Q$ is given by~\cite{Voyles}, 

\be 
\fl 
\langle I({\bf k}, Q) \rangle = \frac{1}{2} \: \pi \: f^2({\bf k}) \: \lambda^2 \: 
\rho_0 \: t \: \left(1 + \rho_0 \: \int d^3{\bf r_{12}} \: g_2({\bf r_{12}}) \: F_k({\bf r_{12}})
a_Q({\bf r_{12}}) \right) 
\ee
where $g_2(r)$ is the radial distribution function, $F_k(r)$ is the coherence function describing 
incoming illumination, and $a_Q(r)$ is the microscope response function. The intensity in the above 
expression involves only $g_2(r)$ and therefore does not carry information about MRO. It is the 
second moment of the intensity $\langle I^2({\bf k}, Q)\rangle$ that includes 3- and 4-body correlation 
functions, which provide information at the medium range length scale.  
A mathematical expression of $\langle I^2(k, Q)\rangle$ and its derivation is given by 
Voyles~\cite{Voyles}.

Computer simulations have recently indicated \cite{Voyles} that amorphous silicon or germanium 
films may contain some nano-sized crystalline grains embedded in a CRN matrix~\cite{Nakhmanson}. 
This model of amorphous silicon is called 
paracrystalline, and simulation of FEM data using these models have been observed to interpret 
experimental results~\cite{Khare, Nakhmanson}. It is proposed that the size and shape of the grains 
are related to the height and position of the peaks in the FEM signal, and an appropriate concentration 
(typically 20\% -- 30\% by number) of such crystalline grains in amorphous matrix can reproduce 
correct structural, vibration and electronic properties~\cite{note3}. However, the model is not unique. Since we know 
from Reverse Monte Carlo simulation that it is possible to generate configurations of amorphous 
silicon having almost identical structure factor observed in experiment but with drastically different 
local bonding, it is necessary to explore the possibility of constructing models that do not explicitly
contain nano-sized grains in the networks to start with. We have studied the problem along this 
direction via reverse Monte Carlo and modified Wooten-Winer-Weaire (WWW)~\cite{www} method and observed 
that direct inclusion of FEM signal in CRN introduces strain in the network~\cite{Ray}. The resulting 
network shows a strong FEM signal and maintains other properties of a-Si, but does not produce any 
visible ordering (such as distorted crystals that is expected from paracrystalline models) except
occasional occurrences of few Schl\"afli clusters~\cite{cluster1, cluster2}.  It is instructive to study 
the stability of paracrystalline models via ECMR. To this end, we first generate a starting configuration 
containing grain(s) of diamond crystal by creating voids of nanometer size in a CRN, and then construct 
a generalized cost function involving FEM signal, a suitably chosen energy functional (modified Stillinger Weber 
potential~\cite{sw}) and the structure factor as follows: 

\bea 
\xi = \lambda \: \Phi_{m-{\rm sw}} + \sum_{i=1}^{3} \beta_i \: \Gamma_i 
\label{penalty}
\eea 

\bea 
\Gamma_1 & = & \sum_j (V_c(k_j) - V_{exp}(k_j))^2  \nonumber \\
\Gamma_2 & = & 1 - \theta(r - r_c)  \nonumber \\
\Gamma_3 & = & \sum_j (S_c(k_j) - S_{exp}(k_j))^2 \nonumber 
\eea 

\begin{figure} 
\begin{center} 
\includegraphics[width=2.4 in, height=2.4 in, angle=0]{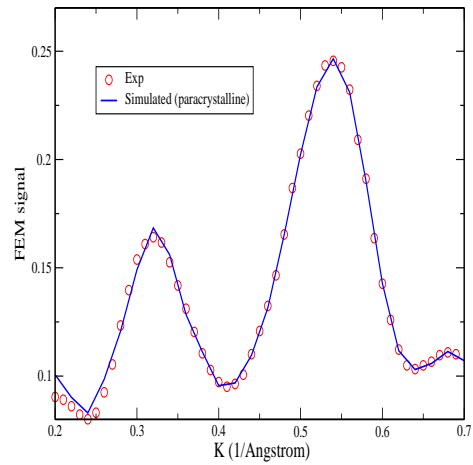}
\caption{ (Color online)
The simulated FEM signal for the final FEM-fitted model (Model-A) along with the experimental 
FEM data. The initial configuration consists of a 216-atom crystal grain in a matrix of 
4056 atoms. The final model is obtained by moving the crystal and interface atoms during 
ECMR minimization.  The experimental signal is multiplied by a factor of 10 in simulation 
and in the plot. 
}
\label{fem} 
\end{center} 
\end{figure}

\begin{figure} 
\begin{center} 
\includegraphics[width=2.4 in, height=2.4 in, angle=0]{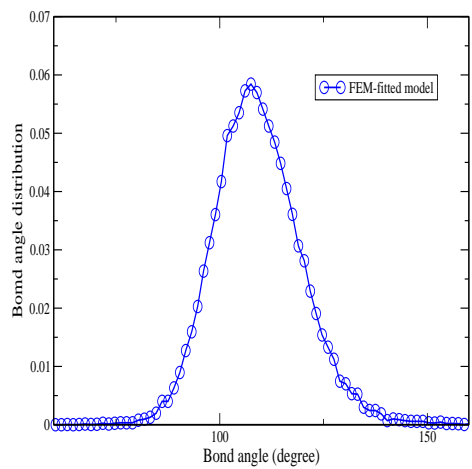}
\caption{ (Color online) Bond angle distribution for the final FEM-fitted model (Model-A) from 
ECMR minimization. The average and root mean square values are $109.7^\circ$ and $11.08^\circ$ respectively. \\ 
}
\label{bond} 
\end{center} 
\end{figure}

\begin{figure} 
\begin{center} 
\includegraphics[width=2.4 in, height=2.4 in, angle=0]{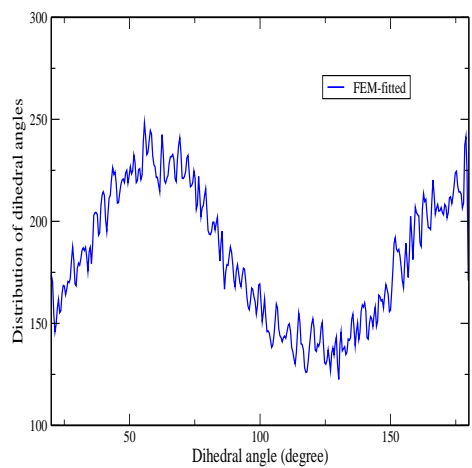}
\caption{ (Color online) Dihedral angle distribution for the FEM-fitted (Model-A) model. }
\label{dihed} 
\end{center} 
\end{figure}

\begin{figure}
\begin{center}
\includegraphics[width=2.4 in, height=2.5 in, angle=0]{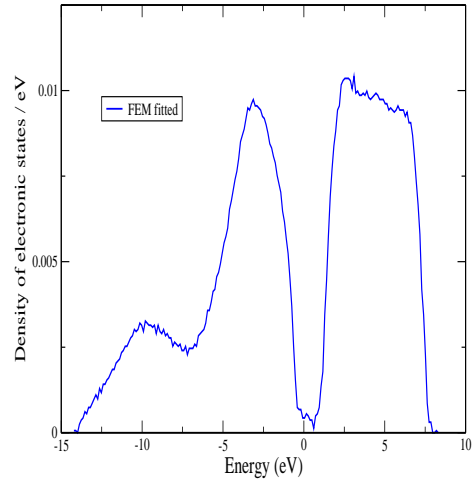}
\caption{ (Color online)
The electronic density of states for the final FEM-fitted model (Model-A) obtained from a 
tight-binding Hamiltonian. A small number of gap states indicate the presence of few 
co-ordination defects in the network. 
}
\label{edos} 
\end{center} 
\end{figure}

Here $\Phi_{m-{\rm sw}}$ is the modified Stillinger-Weber potential, $\Gamma_1$ and $\Gamma_3$ 
stand for FEM data and structure factor respectively, and $\lambda$ and $\beta_i$ are appropriate weight factors 
(for each data set) which may change during the course of simulation.  Our starting configuration is 
a 4056-atom continuous random network that contains a 216-atom grain of diamond crystal. This starting 
configuration shows the presence of a weak FEM signal by construction. We minimize the cost function in equation 
\eref{penalty} via Metropolis Monte Carlo algorithm by moving the crystal and interface atoms~\cite{note1}.  
During the Monte Carlo minimization, the topological constraint of the crystaline grain is relaxed so 
that the atoms in the grain are free to evolve away from (diamond) crystalline geometry, and yet maintain other 
constraints (such as the FEM signal, structure factor etc.). The inclusion of the latter is important 
because of the difference in structure factors of crystalline and amorphous environment of Si. The use 
of modified Stillinger-Weber potential controls the network strain, and maintains the total energy of the system 
during Monte Carlo simulation as minimum as possible. In figure \ref{fem}, we have plotted the simulated  
FEM signal obtained from the final configuration  along with the experimental data. 
A structural analysis of this final configuration shows that the crystal and interface atoms have moved 
significantly to form a distorted ordered structure away from the perfect crystal. A Schl\"{a}fli 
cluster analysis~\cite{cluster1,cluster2} has shown the presence of $ 6_2. 6_2. 6_2. 6_2. 6_2. 6_2 : 29$ cluster 
which originates from diamond crystal structure. The bond and dihedral angle distributions have been plotted in 
figures \ref{bond} and \ref{dihed} respectively.  No significant differences have been observed in the bond 
angle and dihedral distributions compared to its CRN counterpart. The electronic density of states (EDOS) for 
the final FEM-fitted model (Model-A) is plotted in figure \ref{edos} using a tight-binding model Hamiltonian. 
The density of electronic states show a gap with some states in the gap. This is due to the presence 
of few 3-fold and 5-fold coordination defects in the model.

\section{Continuous random network with voids}

A very different approach to understand the FEM signal and hence MRO in amorphous silicon is to study the presence of voids in the network structure. Voids are a universal feature in amorphous 
silicon, and the characteristic of voids depends largely on the growth condition of 
the materials. The presence of voids is considered to be one of reasons of low density 
of amorphous silicon compared to its crystalline counterpart~\cite{Moss}. Small angle scattering of neutrons, 
electrons, and X-rays have been widely used to detect the characteristic presence of voids in both 
amorphous and hydrogenated amorphous silicons~\cite{konnert}. Theoretical modeling of voids in 
amorphous silicon by Biswas {\em et al.} have indicated the presence of rapidly increasing structure factor for 
wave vectors below {1 \AA$^{-1}$}, which is supported by experiments~\cite{rbiswas}.  In this work, we have 
developed models with voids in large continuous random network and have studied the variation of FEM signal with 
different number of voids and its size.

\begin{figure} 
\begin{center} 
\includegraphics[width=2.4 in, height=2.4 in, angle=0]{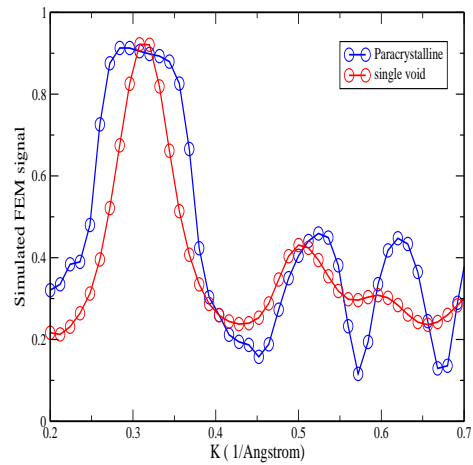} 
\caption{(Color online) 
Simulated FEM signal $ \rm V(k)$ obtained from a 1000-atom paracrystalline model with a 429-atom crystalline 
grain. The FEM signal after removing the grain is also plotted in the figure (indicated as single 
void) for comparison.  
}
\label{para} 
\end{center} 
\end{figure}

\begin{figure} 
\begin{center} 
\includegraphics[width=2.4 in, height=2.4 in, angle=0]{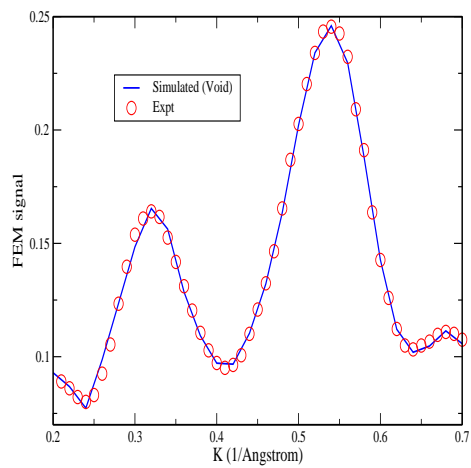} 
\caption{(Color online) Simulated FEM signal $ \rm V(k)$ obtained from a CRN model with a void (Model-B). The experimental data (indicated by circles) 
are used in the model construction via ECMR described in the text. 
}
\label{void} 
\end{center} 
\end{figure}

In order to test the viability of the model, we first start with a 1000-atom paracrystalline model 
and remove the grain of crystal. The resulting model continues to show the presence of FEM signal 
but the strength of the signal decreases as the wave vector increases. In figure \ref{para}, we 
have plotted the FEM signal for a paracrystalline model with and without the crystalline grain. 
It is clear from the figure that the first two peaks have not changed their positions and heights 
significantly. The formation of voids creates some coordination defects and introduces strain in the network, which 
can be minimized by structural relaxation of the network. Using the first-principles density functional 
code {\sc Siesta}~\cite{siesta}, we have relaxed the network to minimize the strain and to reduce the number of 
defects. While the surface of the voids reconstructs, the voids continue to exist in the relaxed model 
with a strong presence of the FEM signal. This observation suggests that presence of voids in amorphous 
network can also produce FEM signal as in paracrystalline model. Together with the presence of 
increasing structure factor at low wave vectors and FEM data, it appears that voids in 
amorphous silicon networks introduce some correlation that can affect the higher order correlation 
functions. Furthermore, introduction of voids does not change the other characteristic material 
properties significantly (such as vibration and electronic density of states). In figure \ref{void} we 
have plotted the results obtained from a model containing 
a single void of radius 12 {\AA}. 
Using our ECMR method, we have minimized the generalised penalty function (equation \eref{penalty}) by moving the 
interface atoms. The void persists, but the surface of the void reconstructs to match with the 
normalized variance of intensity obtained from FEM experiments.

\begin{figure} 
\begin{center} 
\includegraphics[width=2.4 in, height=2.4 in, angle=0]{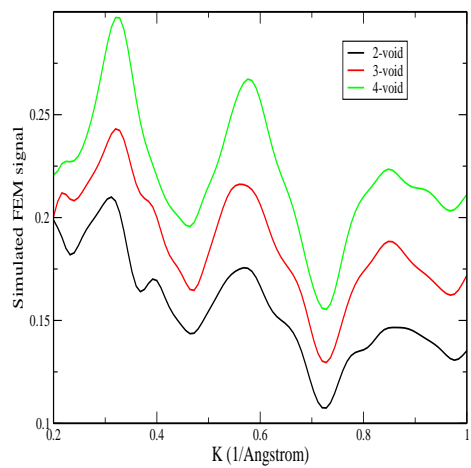} 
\caption{(Color online) 
Simulated FEM signal $ \rm V(k)$ for different number of voids present in a starting 4096-atom 
CRN models. The linear dimensions of the voids are of the order of 6{\AA} to 10 {\AA}.  
}
\label{fig-nvoid} 
\end{center} 
\end{figure}

\begin{figure} 
\begin{center} 
\includegraphics[width=2.4 in, height=2.4 in, angle=0]{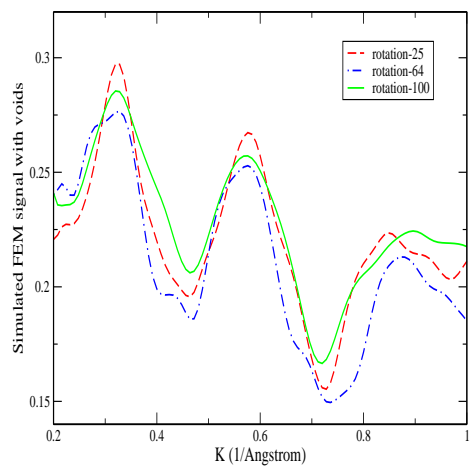} 
\caption{(Color online) 
Simulated FEM signal for different orientation of the model with 4 voids of linear dimension between 
6 {\AA} to 10 {\AA}. The number of orientation is indicated in the figure 
and the average values of the signal are plotted.  }
\label{fig-rvoid} 
\end{center} 
\end{figure}

In figure \ref{fig-nvoid}, we have plotted the simulated FEM signal for different number of voids. 
The signal is observed to be maximum for four voids while minimum for two voids as shown in the figure. 
It is important to note that similar trends have been observed in case of paracrystalline model, 
where signal strength is observed to be dependent on the number of crystalline grains present in the 
sample. We have also studied the role of rotation of the sample for a model with given number of 
voids. The result is shown in the figure \ref{fig-rvoid}. For the model with four voids of linear 
size between 6 {\AA} to 10{\AA}, we find that the signal is more or less independent of 25 to 100 orientations of the model.

\section{Conclusion} 

We have used Fluctuation Electron Microscopy data to incorporate medium range order in 
amorphous silicon starting with continuous random networks. We have discussed two models 
that are capable of producing the characteristic FEM signal observed in experiments 
maintaining structural, electronic and vibrational properties of amorphous silicon. The first model (Model-A) is consists of a CRN with nano-sized ordered grains in the network, while the second model (Model-B) is based on presence of voids in the network. Our study clearly indicates that the FEM signal is sensitive to the presence of small ordered grains and voids in the network. The FEM signal is found to be determined by fluctuations or inhomogeneities due to voids or phase-separated regions of nano-meter size dispersed in approximately homogeneous medium described by continuous random network. We have shown that either crystalline inclusions or voids are possible explanations for the measured FEM data. 

\ack
DAD thanks the US NSF for support under Grants DMR 0605890 and 0600073. PB acknowledges 
the support of the University of Southern Mississippi under Grant No. DE00945. The 
authors would like to thank John Abelson and Paul Voyles for providing experimental 
FEM data, and many conversations and Mike Treacy for the program for Schl\"{a}fli cluster 
analysis.

\newpage 

\newpage 
\begin{figure} 
\begin{center} 
\includegraphics[width=2.2 in, height=2.2 in, angle=0]{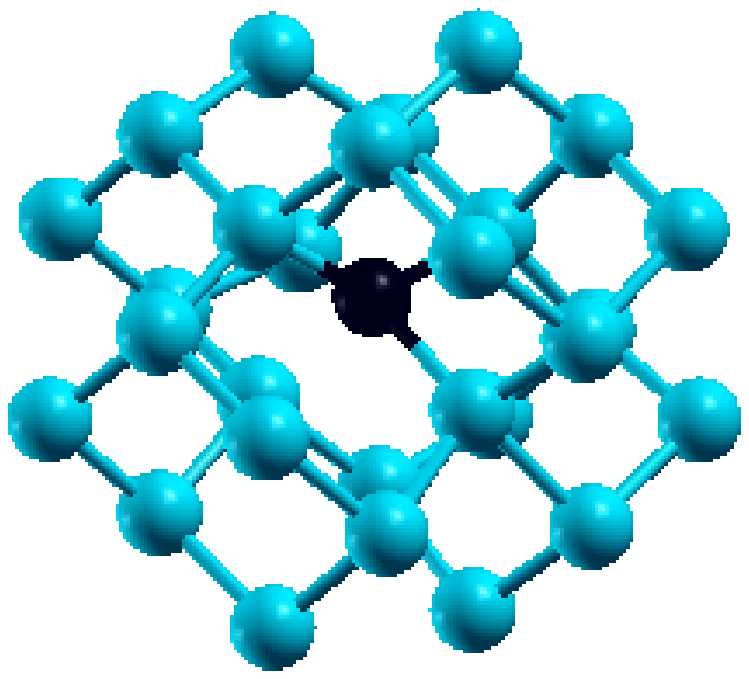} 
\includegraphics[width=2.2 in, height=2.2 in, angle=0]{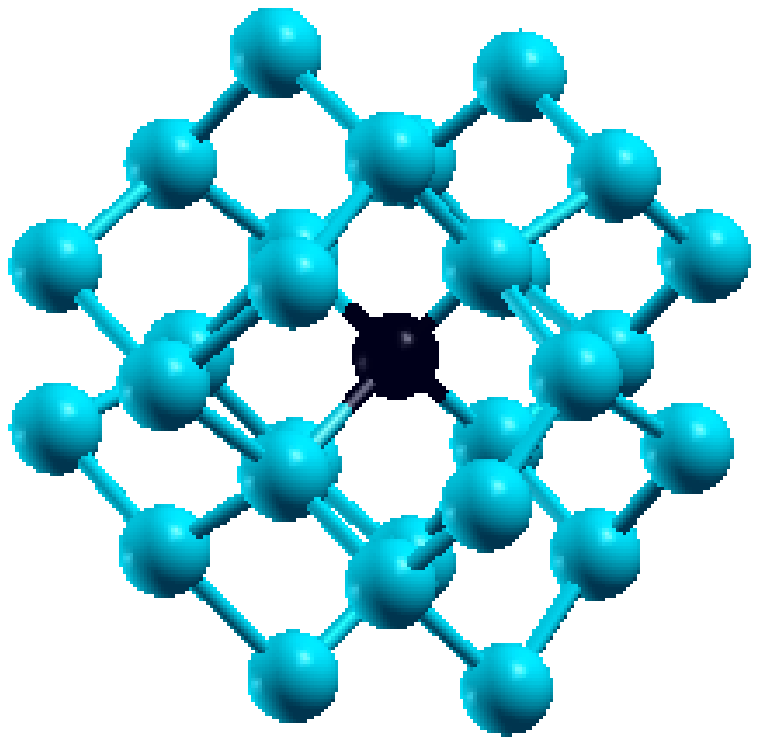} 
\caption{ (Color online) Two representative Schl\"{a}fli clusters $ 6_2. 6_2. 6_2. 6_2. 6_2. 6_2 : 29$ 
found in the FEM-fitted network (Model-A) that originate from diamond crystals. The linear dimension of 
the clusters are 9.1 {\AA} (left) and 9.8 {\AA} (right) respectively (A high quality figure is available 
from the authors on request). 
}
\label{cluster} 
\end{center} 
\end{figure}

\newpage 

\newpage 

\newpage 

\newpage 

\newpage 

\newpage 

\newpage 


\begin{thebibliography}{99}
\bibitem{note3} 
Large crystalline grains in paracrystalline model produce the characteristic 3rd 
crystalline peak in the radial distribution which is absent in amorphous silicon.
This limits the size and volume of the small crystallites present in the network. 
Our work suggests that grains are not crystalline but do contain some topological 
character of (diamond) crystal and is supported by Schl\"{a}fli cluster analysis.
\bibitem{note1}
Computer modeling has indicated that introduction of small grains of crystal always introduce 
fluctuation of intensity measured via normalized variance which is independent of the matrix 
(be it completely disordered, amorphous or otherwise). In our work, we move the crystalline 
and interface atoms with a view to search for configurations that would further 
enhance the FEM signal. This is necessary to produce models compatible with the experimental 
radial distribution function.
\end{thebibliography}
\end{document}